\documentclass[12pt,a4paper,titlepage]{article}
\pdfoutput=1                          
\usepackage[utf8x]{inputenc}
\usepackage[english]{babel}

\usepackage{ucs}
\usepackage{amsmath}
\usepackage{amsfonts}
\usepackage{amssymb}
\usepackage{eucal}
\usepackage{mathrsfs}          
\usepackage[sans]{dsfont}     
\usepackage{slashed}
\usepackage[Symbol]{upgreek}  
\usepackage{array}            
\usepackage{cite}               
\usepackage[normalem]{ulem}             
\usepackage{booktabs}      
\usepackage{colortbl}         
\usepackage{fancyhdr}
\usepackage{multirow}
\usepackage{feynmf}           
\usepackage
{graphicx}

\usepackage{color}
\definecolor{grigio}{cmyk}{0,0,0,0.1}
\definecolor{rosa}{cmyk}{0,0.1,0.1,0.02}
\definecolor{rosino}{cmyk}{0,0.05,0.05,0.02}
\definecolor{rosas}{cmyk}{0,0.3,0.25,0.05}
\definecolor{celeste}{cmyk}{0.1,0,0,0.02}
\definecolor{giallino}{cmyk}{0,0,0.1,0.02}
\definecolor{rosso}{cmyk}{0,1,1,0.4}
\definecolor{rossos}{cmyk}{0,1,1,0.55}
\definecolor{rossoc}{cmyk}{0,1,1,0.2}
\definecolor{blu}{cmyk}{1,1,0,0.3}
\definecolor{blus}{cmyk}{1,1,0,0.5}
\definecolor{bluc}{cmyk}{1,1,0,0.1}
\definecolor{blucc}{cmyk}{0.7,0.5,0,0}
\definecolor{viola}{cmyk}{0,1,0,0.6}
\definecolor{viola2}{cmyk}{0,1,0.2,0.6}
\definecolor{verde}{cmyk}{0.92,0,0.59,0.25}
\definecolor{verdec}{cmyk}{0.92,0,0.59,0.15}
\definecolor{verdes}{cmyk}{0.92,0,0.59,0.4}
\definecolor{verdino}{cmyk}{0.12,0,0.09,0.02}
\definecolor{giallo}{cmyk}{0,0,1,0}
\definecolor{gialloverde}{cmyk}{0.44,0,0.74,0}
\definecolor{Titolo}{rgb}{0.752941176,0.576470588,0.992156863}
\definecolor{altro}{rgb}{0.094117647,0.650980392,0.643137255}
\definecolor{Peanuts}{rgb}{0.2, 0.4, 0.6}
\definecolor{Pean1}{rgb}{0.6, 0.8, 0.4}
\definecolor{BHO}{rgb}{0.2, 0.8, 1}
\definecolor{Daria}{rgb}{0, 0.9412, 0}
\definecolor{UniPi}{rgb}{0.2549, 0.4627, 0.6275}
\definecolor{UniPidue}{rgb}{0.3216, 0.5804, 0.7882}
\definecolor{rossoCP3}{cmyk}{0,.88,.77,.40}
\definecolor{verdeCP3}{rgb}{0.09765625, 0.57421875, 0.1015625}
\definecolor{bluCP3}{rgb}{0, 0.23, 0.67}
\definecolor{bluSaclay}{rgb}{0, 0.22, 0.70}

\usepackage{cancel}
\usepackage[normalem]{ulem}

\newcommand{\eg}{e.g.~}
\newcommand{\ie}{i.e.~}

\newcommand{\Op}{{\cal O}}		
\newcommand{\Mel}{\CMcal{M}}	

\font\bb=bbmss10 scaled 1200
\newcommand{\unop}{\mbox{\bb 1}}
\newcommand{\ud}{\text{d}}
\newcommand{\beq}{\begin{equation}}
\newcommand{\eeq}{\end{equation}}

\newcommand{\Ref}[1]{Ref.~\cite{#1}}           

\newcommand{\ER}{E_\text{R}}

\newcommand{\NR}{\text{NR}}

\def\sfrac#1#2{{\textstyle\frac{#1}{#2}}}

\def\XENONexp{{\sc Xenon}}
\def\CDMS{{\sc Cdms}}
\def\XENON{{\sc Xenon100}}
\def\XENONten{{\sc Xenon10}}
\def\LUX{{\sc Lux}}
\def\FERMI{{\sc Fermi-Lat}}
\def\IceCube{{\sc IceCube}}

\long\def\symbolfootnote[#1]#2{\begingroup\def\thefootnote{\fnsymbol{footnote}}\footnote[#1]{#2}\endgroup}

\ifx\pdfoutput\undefined
\usepackage[dvips,bookmarks]{hyperref}    
\else
\usepackage{hyperref}    
\fi
\hypersetup{colorlinks, bookmarksopen, bookmarksnumbered,
citecolor=verdeCP3, linkcolor=bluCP3, pdfstartview=FitH, urlcolor=rossoCP3}

\def\hhref#1{\href{http://arxiv.org/abs/#1}{#1}} 
\def\mhref#1{\href{mailto:#1}{#1}}        

\begin{document}

\begin{titlepage}

\begin{flushright}
\scriptsize
CCTP-2013-13 \hfill
CP$^3$-Origins-2013-035 DNRF90 \hfill
DIAS-2013-35 \hfill
IPPP/13/75 \hfill
DCPT/13/150
\end{flushright}
\color{black}
\vspace{0.3cm}

\begin{center}
{\LARGE{\color{rossoCP3}\bf Constraints on Majorana Dark Matter \\ from a Fourth Lepton Family }\rule{0pt}{25pt}}
\end{center}

\par \vskip .2in \noindent

\begin{center}
{\sc Tuomas Hapola$\, {\color{bluCP3}^{1,2}}$\,\symbolfootnote[1]{\mhref{t.a.hapola@durham.ac.uk}},
Matti J\"arvinen$\, {\color{bluCP3}^{3}}$\,\symbolfootnote[2]{\mhref{mjarvine@physics.uoc.gr}},
Chris Kouvaris$\, {\color{bluCP3}^2}$\,\symbolfootnote[3]{\mhref{kouvaris@cp3.dias.sdu.dk}}, \\ 
Paolo Panci $\, {\color{bluCP3}^{2,4}}$\,\symbolfootnote[4]{\mhref{panci@cp3-origins.net}},
Jussi Virkaj\"arvi$\, {\color{bluCP3}^2}$\,\symbolfootnote[5]{\mhref{virkajarvi@cp3-origins.net}}}
\end{center}

\begin{center}
\par \vskip .1in \noindent
{\it ${\color{bluCP3}^1} \, $\href{http://www.ippp.dur.ac.uk}{Institute for Particle Physics Phenomenology}, Durham University,\\ South Road, Durham DH1 3LE, UK}
\\
{\it ${\color{bluCP3}^2} \, $\href{http://cp3-origins.dk}{CP$\,^3$-Origins and DIAS}, University of Southern Denmark,\\
Campusvej 55, DK-5230 Odense M, Denmark}
\\
{\it ${\color{bluCP3}^3} \, $\href{http://hep.physics.uoc.gr}{Crete Center for Theoretical Physics}, University of Crete,\\ 
71003 Heraklion, Greece}
\\
{\it ${\color{bluCP3}^4} \, $\href{http://www.iap.fr}{Institut d'Astrophysique}, UMR 7095 CNRS,  \& Universit\'e Pierre et Marie Curie,  \\
98bis Boulevard Arago, FR-75014 Paris, France}

\par \vskip .5in \noindent

\end{center}

\begin{center}
{\large Abstract}
\end{center}

\begin{quote}

We study the possibility of dark matter in the form of heavy neutrinos from a fourth lepton family with helicity suppressed couplings such that dark matter is produced thermally via annihilations in the early Universe. We present all possible constraints for this scenario coming from LHC and collider physics, underground direct detectors, neutrino telescopes, and indirect astrophysical searches. Although we embed the WIMP candidate within a model of composite dynamics, the majority of our results are model independent and applicable to all models where heavy neutrinos with suppressed couplings account for the dark matter abundance. 

\end{quote}

\end{titlepage}

\newpage

\tableofcontents

\newpage

\section{Introduction}

There is a strong possibility that dark matter (DM) might be in the form of weakly interacting massive particles (WIMPs). The most natural candidate within the Standard Model (SM) would have been the neutrinos. However, the existing light neutrinos account for a tiny amount of the matter content of the universe today and in addition light neutrinos would erase large scale structure as they would form ``hot dark matter'' that is not supported by observations. In fact neutrinos lighter than 500 eV would not be able to be packed within a dwarf galaxy due to Pauli blocking~\cite{Tremaine:1979we}. For hypothetical neutrinos heavier than 500 eV, the Weinberg-Lee limit suggests a mass higher than $\sim$3 GeV (or $\sim$13 GeV in the case of Majorana neutrinos) in order not to produce DM higher than the critical density for $\Omega=1$~\cite{Lee:1977ua}. Unfortunately Dirac or Majorana neutrinos with the standard couplings to the other particles are favored neither by collider experiments nor by direct DM 
searches. The width of the invisible decay of the $Z$ boson does not leave any room for Dirac (Majorana) neutrinos, with SM-like couplings, below $\sim45~ (39.5)$ GeV \cite{PDG2012}, and LHC has recently disfavored the existence of a fourth quark family below $\sim400-600$ GeV \cite{Chatrchyan:2012yea,Chatrchyan:2012vu,ATLAS:2012aw,Aad:2012xc}. 
Although this is not directly associated to leptons, it disfavors the neutrino DM scenario since it would be strange to have a fourth lepton family not accompanied by a fourth quark family. In addition in such a case, Witten's $SU(2)$ global anomaly will not be cancelled. On the other hand direct search experiments such as \CDMS\ and \XENONexp\ have imposed strong bounds on the mass of heavy SM neutrinos. For Dirac SM neutrinos the bound is tens of TeV, whereas for Majorana ones, the strictest bound from \XENONexp\ is $\sim 2$ TeV~\cite{Angle:2008we}.

Although SM neutrinos have been more or less abandoned as a solution to the DM problem, non-standard neutrinos are still good candidates. One typical example is the case of sterile neutrinos~\cite{Dodelson:1993je,Shi:1998km,Dolgov:2000ew,Gorbunov:2007ak,Shaposhnikov:2008pf,Laine:2008pg,Bezrukov:2008ut} that are still viable DM candidates. Another type of candidates arises from the observation that heavy neutrinos that have different (suppressed) couplings from the SM ones, can 
also have the appropriate annihilation cross section in the early universe epoch to be produced thermally. An extra benefit for such candidates is the fact that the suppression of couplings corresponds generally to lower WIMP-nucleon cross sections, thus lowering the bounds on the neutrino masses from direct search experiments. This suppression of the couplings can be implemented in the context where a left-handed neutrino mixes with a right-handed one producing two Majorana states. If the lightest among them is mostly right-handed, it can constitute the DM. This is a sort of inverse seesaw-like mechanism, in which the amount of mixing between the left and right-handed neutrinos determines the size of the couplings to the $Z$ boson and Higgs and consequently controls the amount of annihilation in the early universe~\cite{Enqvist:1988nd,Enqvist:1988dt}. 
Notice that if the usual see-saw mechanism is used in the present heavy neutrino set up, the lightest particle becomes dominantly left handed and has SM-like couplings to Higgs and SM gauge bosons. As we mentioned this scenario is ruled out by present experimental data (LEP, \XENONexp ).

This idea of  suppressed couplings of Majorana neutrino type particles can be nicely accommodated in Technicolor (TC) theories. Although TC provides naturally asymmetric type of dark matter~\cite{Nussinov:1985xr,Barr:1990ca,Gudnason:2006ug,Gudnason:2006yj,Khlopov:2007ic,Kouvaris:2008hc,Foadi:2008qv,Khlopov:2008ty,Sannino:2008wy,Ryttov:2008xe,Frandsen:2009mi}, and mixed DM candidates (with a thermal and a nonthermal component)~\cite{Belyaev:2010kp}, purely thermally produced DM candidates can be easily facilitated~\cite{Kainulainen:2006wq,Kouvaris:2007iq,Kainulainen:2009rb} even with a possibility to enlarge the DM model to provide the unification of SM gauge coupling constants~\cite{Kainulainen:2010pk,Kainulainen:2013sva}. In such TC models the 125 GeV Higgs may be realized as a light scalar composite state, (see  \eg \cite{Foadi:2012bb} and references therein). 

There are different TC frameworks upon which the suppression scenario can be visualized. In this article we are going to concentrate on the Minimal Walking Technicolor (MWT)~\cite{Sannino:2004qp,Dietrich:2006cm,Gudnason:2006ug,Gudnason:2006yj,Foadi:2007ue}. However the heavy neutrino DM scenario  can be identically realized for example in a partially gauged TC model as we shall explain later (partially gauged TC models have been considered \eg in  \cite {Dietrich:2005jn, Christensen:2005cb, Dietrich:2006cm, Dietrich:2008ni, Dietrich:2008up,Luty:2008vs,Ryttov:2008xe,Kainulainen:2013sva}). There are also more involved Extended TC (ETC) models~\cite{Appelquist:2003uu} in which this kind of DM scenario could exist~\cite{Appelquist:2003hn}. 

  The particle content of the MWT theory is simple. It postulates two flavors of techniquarks $U$ and $D$ that transform under the adjoint representation of an $SU(2)$  gauge group. Since techniquarks couple to the electroweak (EW) sector,  in order to cancel Witten's global anomaly,  an extra lepton family is included. As we shall describe in detail in the next section, gauge anomalies are canceled with a proper choice of hypercharge assignment for techniquarks and new leptons (that is not unique). The helicity suppression scenario can be accommodated within MWT in two ways. 

In the first case, for an appropriate choice of hypercharge, $D$ techniquark (for example) becomes electrically neutral. Due to the fact that $D$ as well as technigluons transform in the adjoint representation of the $SU(2)$ group, composite states of the form $D_LG$ and $D_RG$ (where $L(R)$ denote left (right) particles and 
$G$ technigluons) are not only electrically neutral but also colorless. In terms of quantum numbers $D_LG$ and $D_RG$ behave as left and right handed neutrinos. Using the inverse sewsaw-like mechanism, \ie introducing apart from a Dirac mass, Majorana masses for the left and right composite objects $DG$, leads to the creation of two states with suppressed couplings to the $Z$ boson~\cite{Kouvaris:2007iq,Kouvaris:2008hc}. The lightest of the two particles is the DM. The mixing between left and right particles can be adjusted so it gives the correct thermal abundance of DM. One should note that for the hypercharge assignment that makes $D$ neutral, none of the new leptons is electrically neutral. They have charges $-1$ and $-2$ respectively. In the second case, one can choose a SM-like hypercharge assignment that makes one of the leptons electrically neutral~\cite{Kainulainen:2006wq,Kouvaris:2007iq,Kainulainen:2009rb}. Similarly as before, one can introduce Majorana mass terms for the new left and right handed 
electrically neutral leptons. By adjusting the masses, two Majorana mass eigenstates result, where again the lightest one is the DM. 
Naturally, we assume here that there is no mixing between the fourth and the SM lepton families and that the lightest lepton from the fourth family is stable (for example due to a conserved quantum number).
Once more, the couplings to the $Z$ and the Higgs are helicity suppressed, and can be adjusted in such a way that the observed DM relic density is produced.  

In this paper we are going to focus on the second scenario, \ie the case where the new lepton family provides an electrically neutral particle that will be the DM WIMP. However, most of our analysis and results can be directly interpreted in the light of the $DG$ DM scenario. For example, the relic density analysis and the DM direct detection constraints are practically identical in both models. Furthermore, our analysis for the fourth lepton family DM can be also considered out of the MWT context,  \ie as an independent study of a fourth family heavy neutrino DM. Although in this case our EW precision test analysis is not directly applicable, our estimates of the relic abundance and the constraints are independent of the details of the theory the fourth lepton family is embedded\footnote{We can consider for example a partially gauged TC model, in which the TC gauge group is SU(3) and the techniquarks transforms under the fundamental presentation of the TC group, but only one doublet of techniquarks is 
charged under electroweak gauge group. 
Then the EW gauged TC quark doublet, when accompanied by a new lepton doublet, with SM-like hypercharge, appears just like the fourth SM family, and the model is free of anomalies. Furthermore, the EW precision test analysis is identical to the  MWT one, as the TC sector gives the same contribution to the S parameter in both cases. In addition, this model does not have the potential problems related to the $DG$ states, appearing in the MWT, as we shall discuss later in the text.}.
Thus our results are generally valid for any model where DM is in the form of  a thermally produced heavy neutrino with suppressed couplings to $Z$ and Higgs bosons.

We examine the fourth heavy neutrino DM scenario from every different perspective. 
We identify what phase space of the theory produces the correct DM relic density and avoids the tight constraints of direct search experiments and EW precision tests. We study the annihilation of this DM candidate to all possible channels and we analyze the constraints derived from indirect detection.
We analyze the possibility of detecting our DM candidate at LHC. Although general constraints from monophoton and monojet processes, based on an effective field theory approach, have been already presented in the literature, these constraints rely on the assumption that the WIMP-nucleon interaction is mediated by a heavy particle (contact interaction). However in our case, since the production of the WIMP is mediated by either the $Z$ or the Higgs, the existing constraints in literature are not valid and an analysis from scratch is needed. 
We combine, improve and update the analysis done previously in~\cite{Kouvaris:2007iq,Kainulainen:2009rb,Heikinheimo:2012yd,Belotsky:2008vh}, and  
in particular we study and present several new constraints for this model.

The paper is organized as follows: We present the model in the next section. The model analysis and constraints are described in Sec.~\ref{sec:analysis}. In particular, the relic density analysis and the constraints from the EW precision tests are presented in Sec.~\ref{sec:omega} and the other collider constraints are discussed in Sec.~\ref{sec:collider}. The direct and indirect constraints are presented in Sec.~\ref{sec:direct} and Sec.~\ref{sec:indirect} respectively. Finally we present our results and conclude in Sec.~\ref{sec:results}.

\section{Model}
\label{sec:model}

As we mentioned in the introduction, MWT has two techniquarks $U$ and $D$ transforming under the adjoint representation of an $SU(2)$ TC gauge group. The left-handed $U$ and $D$ form a doublet that is gauged under the $SU(2)$  EW gauge group (as it happens in all conventional TC models). As in the case of quarks, right-handed ones do not couple to the EW $SU(2)$. Due to the fact that $U$ and $D$ transform under the adjoint of the TC group, there is an enhanced $SU(4)$ global symmetry that includes as a subgroup the $SU(2)_L \bigotimes SU(2)_R$. The chiral symmetry breaking pattern is $SU(4)\rightarrow SO(4)$, where $SO(4)$ contains as a subgroup the $SU(2)_{L=R}$. There are nine Goldstone bosons arising from the symmetry breaking. Three of them, eaten by the EW-gauge boson, are pion-like; their composition is that of pions made of $U$ and $D$. There are also three Goldstone bosons of di-quark type $UU$, $UD$, and $DD$ and their antiparticles. These particles carry a conserved technibaryon quantum number.  
Notice that the exact form of these Goldstone bosons is $Q^{aT}CQ^a$ where $Q$ is either $U$ or $D$, $C$ is the charge conjugate matrix, and $a$ is the three states of technicolor in the adjoint representation ($rr$, $rg+gr$, $gg$), where $r$ and $g$ are the two technicolors. 
The EW symmetry breaks due to the formation of condensates $\bar{q}_Lq_R$ (and h.c.) and the Higgs boson is composite and of the form $\bar{U}U+ \bar{D}D$. However, since the $(U,D)$ doublets transform under the adjoint of the TC $SU(2)$, there are three new doublets introduced. In order to avoid Witten's global anomaly \cite{Witten:fp}, there is a need for an extra lepton family. The doublet  $L_L=(N_L, E_L)$ couples to the EW $SU(2)$ making consequently even the total number of doublets (avoiding thus the anomaly).

Apart from Witten's global anomaly, one should in principle worry about gauge anomalies. All gauge anomalies are canceled if a proper hypercharge assignment is given to the new particles. The only gauge anomalous free choice is \cite{Dietrich:2005jn}
\beq
Y(Q_L)=\frac{y}{2},\quad Y(U_R,D_R)=\left(\frac{y+1}{2},\frac{y-1}{2}\right), \label{hyp_q}
\eeq for the techniquarks,
and 
\beq
Y(L_L)=-\frac{3y}{2},\quad Y(N_R,E_R)=\left(\frac{-3y+1}{2},\frac{-3y-1}{2}\right). \label{hyp_l}
\eeq for the new leptons. Gauge anomalies cancel for any real value of the parameter $y$. The choice $y=1$ renders the $D$ techniquark electrically neutral, $U$ with charge $+1$ while the new leptons get charges $-1$ and $-2$. The Standard Model-like choice $y=1/3$, makes $U$ and $D$ having charges $+2/3$ and $-1/3$ respectively, while $N$ and $E$ have charges 0 and $-1$. From this point of view, $N$ appears as a new (heavy) neutrino bearing the quantum numbers of the usual neutrinos. 

\subsection{Lagrangian and mass terms}
\label{sec:Lagrange}

The full Lagrangian of the theory is
\begin{eqnarray}
\mathcal{L} & = & \bar{Q}_L\gamma^{\mu}D_{q\mu}Q_L +\bar{L}_L\gamma^{\mu}D_{l\mu}L_L +
  \sum_i \bar{Q}_{Ri}\gamma^{\mu}D^q_{i\mu}Q_{Ri} + \sum_i \bar{L}_{Ri}\gamma^{\mu}D^l_{i\mu}L_{Ri}  \nonumber \\ 
 & +  &  \mathcal{L}_{\text{mass}} + \mathcal{L}_{SM}, \label{lag}
\end{eqnarray}
where the sum $Q_{Ri}$ is over $U_R$ and $D_R$, and the sum $L_{Ri}$ is over $N_R$ and $E_R$. The EW covariant derivatives are presented by $D_{q\mu}$ and $D_{l\mu}$ (keeping in mind that the hypercharge assignments for left techniquarks $q$ and left leptons from the fourth family $l$ are different). Similarly $D^q_{i\mu}$ and $D^l_{i\mu}$ represent the covariant derivatives for the right-handed techniquarks and fourth family leptons respectively. Note that right-handed particles couple only to the hypercharge $U(1)$ as shown in Eqs.~(\ref{hyp_q}) and (\ref{hyp_l}).  $ \mathcal{L}_{SM}$ is the Standard Model Lagrangian  and  $ \mathcal{L}_{\text{mass}}$ represents 
all possible mass term sources for the new leptons. It is given by
\begin{eqnarray}
{\mathcal{L}}_{{\rm{mass}}} 
  &=& (y_E \bar{L}_L H E_R+ {\rm{h.c.}})+C_D\bar{L}_L\tilde{H}N_{R}
\nonumber \\
  &+& \frac{C_{L}}{\Lambda}(\bar{L_L}\tilde{H})(\tilde{H}^TL_L^c)
   +  C_{R}S\bar{N}^c_{R}N_{R} 
   +{\rm{h.c.}}
\label{scalar_fermion2}
\end{eqnarray}
where $\tilde{H}=i\tau^2H^\ast$ with $H$ being the SM-like (composite) Higgs doublet.
The parameter $\Lambda$ is an energy scale that is associated with the UV completion of the theory at higher scales. This can be the scale of the ETC or even the scale of Grand Unification (GUT). 
In dynamical EW symmetry breaking the Higgs doublet will get a vacuum expectation value, like in SM, $H \rightarrow (v+h)/\sqrt2$. In this case the first two terms give Dirac masses for $E$ and $N$, whereas the last two terms represent Majorana masses for the left and right-handed $N$ respectively. All terms are gauge invariant. The Majorana mass term for the left-handed $N$ is the usual dimension five Weinberg term. 
Since the operator $\bar{N}^c_{R}N_{R}$ is gauge invariant, there is the possibility that  the right-handed $N$ receives a ``hard'' Majorana mass term by coupling this operator to another gauge invariant scalar operator from an additional sector. We have implemented this possibility in term of a coupling to a new scalar $S$ in Eq.~(\ref{scalar_fermion2}).  
Once $S$ gets a vacuum expectation value, the Majorana mass becomes $M_R=C_Rv_s$ where $v_s$ is the vev of $S$. It should be noted here, that for the right-handed $N$, the Majorana mass could also come from a coupling to the Higgs boson of the form $(C_{R}/\Lambda)(H^\dagger H)\bar{N}^c_{R}N_{R}$ (see \eg \cite{Kainulainen:2009rb}). Although in 
principle such a term gives a Majorana mass similar to the one given by $S$, it is experimentally excluded~\cite{Heikinheimo:2012yd} because once we go to the mass diagonal basis, the coupling of the DM particle (\ie the lightest between the two Majorana particles) to the Higgs boson is always sufficiently large to be excluded by direct DM search experiments. 

 We should also note here that  although TC is sufficient to break the EW symmetry dynamically and can provide the Higgs boson, it does not provide masses to the Standard Model particles per se. All the Yukawa couplings are provided by the ETC interactions. Once the techniquarks form a chiral condensate at the EW scale, these interactions take the form of an effective mass for the Standard Model particles. Constructing a working ETC model is a very difficult task. The ETC scale should be a few hundred times (or more) higher than the EW (although realistically it can take place in three different distinct energy scales, one for each family of the SM). In this paper we are not going to speculate on the specifics of the ETC model. $\Lambda$ can be the ETC scale, or it could be the GUT scale, if the fourth lepton family unifies with the techniquarks at that scale. For our phenomenological study, the details of the ETC and GUT unification are not important. Note 
here that for this reason we have omitted the ETC interactions from our Lagrangian (\ref{lag}). 
 
Before continuing with the model details, let us comment on some issues which might follow from our hypercharge convention. Previously~\cite{Hapola:2012wi} it was pointed out, that the choice of SM-like hypercharges for the new leptons and techniquarks, would make states like $D_LG$ and $D_RG$ fractionally charged. If these are the lightest states in the technibaryon number preserving particle spectrum, they would be stable, and problematic from  a cosmological perspective, as no fractionally charged relics have been observed so far. Some of these aspects were afterwards discussed in~\cite{ Heikinheimo:2012yd}. However here we take a slightly different approach to this issue, as we do not mind if these states are formed in the early universe. Indeed, as we consider the new neutrino to be the DM particle, we may allow the violation of the technibaryon number by ETC interactions. This would imply that the TC particles would not contribute to the DM density today as they could decay by ETC 
interactions to SM particles. As said previously, we do not intend to build a complete ETC model here, but we give a short description how ETC interactions could  get rid of these fractionally charged $D_LG$-like states.
Generally, if the techniquarks and ordinary quarks belong to a common ETC multiplet, after the breaking of the ETC symmetry to TC and QCD at some high energy scale $\Lambda_{\rm{ ETC}}$, the $D_LG$ like states could decay to SM quarks by emitting an ETC gauge boson. Naturally this process would be suppressed by factors of $1/ \Lambda_{\rm{ ETC}}$, but as long as the decay time would be much shorter than the age of the universe at recombination, all the $D_LG$ states formed in the early universe, would have been able to decay by now, leaving no problematic fractionally charged relics at present\footnote{Notice that this kind of problems do not exist in the partially gauged TC model, discussed in the previous footnote, as the formation of the $D_LG$-like state is forbidden by the TC gauge symmetry.}.

To be more quantitative, we can try to make a rough estimate for the DG decay width $\Gamma_{DG} $, to see what could be the lifetime of these particles. 
First very naive attempt could be, to use a simple estimate of a free D techniquark decay to SM quark and ETC gauge boson (similar to top quark decay), but as the DG is a bound state, this would probably give too optimistic result i.e.~too short lifetime for DG.  
However, motivated by QCD meson decay, we can write another estimate $\Gamma_{DG} \sim (g^2_{ETC}/M^2_{ETC})^2 f^2_{DG}  m_{DG} m^2_{t,b}$. Here $g_{ETC}$ and $M_{ETC}$ are the ETC gauge coupling and the ETC gauge boson mass respectively, $f_{DG}$ and $m_{DG}$ are the decay constant and the mass of the DG state respectively, and finally  $m_{t,b}$ refers to top or bottom quark masses, which are used respectively, depending whether $U$ or $D$ of the techniquarks is actually the lighter one. Now using reasonable assumptions for the mass scales of the unknown parameters $M_{ETC}/g_{ETC} \sim \Lambda_{ETC} \sim 10$ TeV  and $m_{DG}\sim  f_{DG} \sim \Lambda_{TC} \sim 1$ TeV,  we get an estimate for the lifetime of the DG states: $\tau_{DG} = 1/\Gamma_{DG} \sim 10^{-22} (10^{-19})$ s, using top(bottom) mass in $\Gamma_{DG}$. This result indicates that these particles could actually decay already well before BBN. However, a more detailed analysis of the evolution (i.e.~decay/freeze-out) of these particles and their possible effects to early universe physics are left for further work.

\subsubsection{Mass mixing pattern}
\label{sec:mass}

Once the TC chiral condensate forms (at the EW scale), and the composite Higgs boson gets a vacuum expectation value, Eq.~(\ref{scalar_fermion2}) provides the mass for the fourth family leptons. Focussing on the neutral one, 
this gives,
\begin{equation}  
   -\frac{1}{2}\bar{n}_L^c 
   \left(\begin{array}{cc} M_L & m_D \\ m_D &  M_R\end{array}\right) n_L
   + h.c. \,,
\label{eq:massmatrix1}
\end{equation}
where $n_L=(N_{L}, N_{R}^{~c})^T$, $m_D=C_Dv/\sqrt{2}$, $M_{L}=C_{L}v^2/2\Lambda$, and $M_R=C_Rv_s$.
It is easily seen that the case $M_L=M_R=0$ corresponds to a Dirac particle, whereas $m_D=0$ corresponds to pure Majorana states for $N_L$ and $N_R$. In the generic case where all three masses ($m_D$, $M_{L,R}$) are nonzero, the mass eigenstates of the system can be given as a linear combination of the gauge eigenstates
\begin{equation}
  N = O n_L + \rho O^T n_L^c \,,
\label{Neigen}
\end{equation}
where $N \equiv (N_1,N_2)^T$ and $O$ is an orthogonal $2\times 2$ rotation matrix, parametrized by a rotation (mixing) angle $\theta$, that is related to the parameters of the matrix (\ref{eq:massmatrix1}) as
\begin{equation}
  \tan 2\theta = \frac{2m_D}{M_R-M_L} \,.
\label{eq:tan}
\end{equation}
The two eigenvalues of the mass matrix of Eq.~(\ref{eq:massmatrix1}) are
\begin{equation}
  M_{1,2}=\frac{1}{2}\Big(M_L+M_R\pm\sqrt{(M_L-M_R)^2+4m_D^2} \;\Big) \,.
\label{eq:masseigen}
\end{equation}
However, although $M_1$ is always positive (in case $M_{L,R}$ are positive too), this is not always the case for $M_2$. The positiveness of both $M_1$ and $M_2$ is ensured by the diagonal matrix $\rho = {\rm diag}(\rho_1,\rho_2)={\rm diag} (\rm{sgn} (M_1), \rm{sgn} (M_2))$. It is clear from the above that $\rho_1 = 1$. It is also not hard to show that $\rho_2=-1$ if $m_D^2>M_LM_R$. 
Although there is phase space for $\rho_2=1$, the phase space of the model presented here is in the former case. 
We checked that a change to the $\rho_2=1$ part of the phase space has very little impact on our final results, i.e. the $\sin \theta$ values. Furthermore, the $\rho_2=1$ phase space is more constrained by the EW precision data\footnote{Notice that in Section 2.3 and in Figs.~1-4 of~\cite{Heikinheimo:2012yd}, there is a typo/mix-up in the $\rho_{12}$  (notation $\rho_{12}=\rho_{1}\rho_{2}$)  values: in each point where $\rho_{12}$ appears one should replace $\rho_{12}=+1$ with $\rho_{12}=-1$ and vice versa.}. This is also why we concentrate on the $\rho_2=-1$ case here.
Note that the pure Dirac scenario ($M_L=M_R=0$), or the pure Majorana one ($m_D=M_R=0$) lie on the $\rho_2=-1$ phase space. A thorough study of this $\rho_2= \pm1$ issue has been presented in~\cite{Kainulainen:2009rb}.

\subsection{WIMP interactions}
\label{sec:int}

The ingredients needed for the calculation of the relic density and the cross section between WIMPs and nucleons are encoded in the couplings of $N$ with the $Z$ and the Higgs boson, as well as in the coupling among $E$, $N$, and $W$. These can be rewritten from the gauge basis to the mass basis as
\begin{eqnarray}\label{eq:intlagrange}
\mathcal{L_{\rm{inter}}}&=&\frac{g}{\sqrt2}W^+_\mu\bar{N}_L\gamma^\mu E_L +\frac{g}{2 \cos \theta_{\rm W}}Z_\mu\bar{N}_L\gamma^\mu N_L =
\frac{g}{\sqrt2} \sin\theta \; W^+_\mu \bar{N}_{2}\gamma^\mu E_L + \\
&+&  \frac{g}{2 \cos \theta_{\rm W}} ( \sfrac{1}{2}\sin^2\theta \; Z_\mu \bar{N}_{2}\gamma^5 \gamma^\mu N_{2} 
+\sfrac{1}{2}\sin 2\theta \;  Z_\mu \,  \bar{N}_{1} \gamma^\mu N_{2} ) \,+ \dots ,  \nonumber                         
\end{eqnarray}
%
%
where $g$ is the weak coupling constant and $\theta_{\rm W}$ is the Weinberg angle. We have omitted interaction terms involving only the heavy $N_1$ field. 

Finally the interaction with the Higgs boson can be read off Eq.~(\ref{scalar_fermion2}) and transformed in the basis of mass eigenstates as
\begin{eqnarray}\label{higgs_interactions}
{\mathcal{L}}_{\rm{Higgs}}  &=& 
- \frac{gM_2}{2M_W}\Big( \; C_{22}^h h\bar{N_2}N_2+C_{21}^h h\bar{N_1}\gamma^5N_2   + \frac{1}{v}C_{22}^{h^2} h^2\bar{N_2}N_2 \,\Big)  \,+ \dots ,
\end{eqnarray}
where $C_{22}^h=\sin^2\theta$,
$C_{21}^h=-\frac{1}{4}\sin2\theta R_+$, $C_{22}^{h^2}=\frac{1}{2}\sin^2\theta (1-\cos^2\theta R_-)$, and $R_{\pm}=1\mp M_1/M_2$. Here we have also omitted the interaction terms involving only $N_1$ as they are irrelevant for our analysis.

\section{Model analysis and constraints}
\label{sec:analysis}
 
We start this section by describing our method of scanning and constraining the parameter space of the model using the EW precision data. 
Then we briefly review the WIMP annihilation of our DM candidate and the parameter phase space where it can account for the DM relic abundance.
We require that the model satisfies the EW precision tests on all the points of the parameter space where the DM relic density is obtained by the candidate.  Finally we present all other relevant experimental constraints coming from colliders, direct and indirect DM detectors. 

\subsection{Model parameter scan and relic density}
\label{sec:omega}

Our strategy is to scan the model parameter space (\ie $m_2$, $\theta$, $m_E$, ($m_1$)) and search for parameters sets that are consistent with the EW precision data. We scan over a large range of WIMP mass ($m_2$ from 20 GeV up to 1 TeV), forcing of course $m_2<(m_1,m_E)$, and the mass of the particles to be smaller than $4\pi \,v \simeq 3$ TeV. Furthermore, in order to avoid coannihilation in the early Universe \cite{Griest:1990kh} and $N_2 \leftrightarrow N_1$ oscillation during the freezeout if an asymmetric DM sector is taken into account \cite{Cirelli:2011ac, Tulin:2012re, Buckley:2011ye}, we always consider a mass gap between $m_1$ and $m_2$ of at least 50 GeV. To reduce the parameter space further we fixed $m_1$ in the scan (we make three different scans using three fixed values $m_1$ = 0.5, 1  and 1.5 TeV). In this set up, it is then the $m_E$, which is mostly restricted by the EW precision data, as the values of $m_2$ and $\sin \theta$ are arranged to produce the correct DM relic density, as we 
shall explain below. 

To compare our model against the EW precision data we use the usual analysis in terms of the oblique parameters $S$ and $T$  \cite{Peskin:1990zt,Peskin:1991sw}. These parameters measure the modifications of the SM gauge boson vacuum polarization amplitudes by the contributions following from the new physics. In our case, the S and T parameters are affected by the new leptons $N_1$, $N_2$ and $E$,  and by the TC sector  \ie by the techniquarks. Our calculation of S and T follow the analysis performed in \cite{Antipin:2009ks} for this model. The experimental values $(\rm{S} ,\rm{T})=(0.04\pm0.09,0.07\pm0.08)$ including 88$\%$ correlation between S and T were taken from\cite{PDG2012}\footnote{These experimental values assume the Higgs mass to be in the range 115.5 GeV $< m_h < $ 127 GeV, which is in agreement with the observed value $m_h \approx 125$ GeV. }. 
We scan the model parameter space (\ie $m_2$, $\theta$, $m_E$, ($m_1$)) and search for parameter sets that render the theory within the experimental 90\% CL (S,T)-contour ellipse. Having in our  disposal the parameter space that satisfies the EW precision data, we then compute the DM relic abundance numerically.

The DM relic density calculation for this particular model has been previously studied in detail in \cite{Kainulainen:2009rb}, and we follow that analysis here. 
The DM relic density is calculated numerically using the usual DM Boltzmann equation.
In the calculation we have assumed the standard radiation driven expansion history of the universe, and
for the thermally averaged annihilation cross section $\langle \sigma v \rangle$ we use the standard integral form given in \cite{Gondolo:1990dk}. Here in the annihilation cross section $\sigma$ we have included all the allowed annihilation channels, \ie process
${N_2}  \bar{N_2} \rightarrow f \bar{f},W^+ W^-, ZZ, Zh$ and $hh$
\footnote{For the annihilation cross section in the case of $N_2 \bar{N_2} \rightarrow  f \bar{f}$ we used an improved version of the expression given in the Appendix of \cite{Heikinheimo:2012yd} including finite width approximation for the resummed Z boson propagator, which captures the gauge invariance correctly \cite{Nowakowski:1993iu,Stuart:1991xk,Baur:1995aa}. The other cross sections for processes $N_2 \bar{N_2} \rightarrow W^+ W^-, ZZ, Zh$ and $hh$ follow from \cite{Kainulainen:2009rb} .
}, where $f$ refers to all SM fermions. We have considered only tree level processes in our analysis. In the s-channel processes the interaction mediator is either the $Z$ or the Higgs boson. In the $t$- and $u$-channels, the mediator, depending on the process, is either $N_2, N_1$ or the new ``electron" $E$. 

As it is well known for symmetric WIMP dark matter, the relic density is essentially dictated by the annihilation cross section. In our case the annihilation cross section is ultimately controlled by the WIMP mass and especially the mixing angle $\sin \theta$. Indeed, all the couplings between $N_2$ and SM particles include a mixing angle dependent factor.
As it can be seen from Eqs.~(\ref{eq:intlagrange}) and (\ref{higgs_interactions}), the couplings of $N_2$ to the $Z$ and Higgs bosons are suppressed relatively to the couplings of a standard Majorana neutrino with no mixing by a factor of $\sin^2 \theta$.  In order to produce the DM relic density, for a given WIMP mass, one should adjust the mixing angle accordingly so the annihilation cross section is the one required for thermal production.
Thus for example, in the vicinity of resonances, like near the s-channel $Z$ or Higgs boson resonances (when the WIMP mass is $m_2 \approx m_Z/2$ or $m_2 \approx m_h/2$), the increase in the annihilation cross section is counterbalanced by a decrease in the mixing angle, so the overall cross section stays constant and equal to the usual thermal value. The same happens when new annihilation channels open up: to compensate the increase of the cross section the mixing angle needs to be suppressed to obtain the right DM density. 

To summarize, for a given WIMP mass $m_2$ and fixed $m_1$, the new fourth heavy ``electron'' mass $m_E$ is determined by demanding that the theory passes the EW precision tests, and the mixing angle $\theta$ is basically set by requiring a correct DM relic production.

In Fig.~\ref{fig:Results} we show our results in the $(m_2,\sin \theta)$-plane, where the black solid, dashed and dotted lines refer to three different fixed $m_1$ values: (0.5, 1, 1.5) TeV producing the right relic density.  The two dips in the $\sin \theta$ values around $m_2 \approx m_Z/2$ and $m_2 \approx m_h/2$ indicate respectively the $Z$ and Higgs bosons resonances. The overall drop in the $\sin \theta$ after $m_2 \ge m_W$ is due to 
the opening of the dominant $N_2 \bar{N_2} \rightarrow  W^+ W^-$ annihilation channel for heavy WIMP masses. The smaller drops in the mixing angle values around $m_2 \approx 91, 108, 125$, and 175 GeV  indicate the openings of new channels  \ie $N_2 \bar{N_2} \rightarrow ZZ, Zh, hh,$ and $t \bar{t}$ respectively. 

Obviously the mixing angle $\theta$ and the WIMP mass $m_2$ set basically the line along which the right DM relic density is obtained. 
This is due to the fact that the dominant  tree level s-channel processes involved only SM mediators, such as either the $Z$ or the Higgs boson. The residual dependence on $m_1$ and $m_E$ is controlled by the t- and u-channels, in which the mediator is either $N_2, N_1$ or the new heavy ``electron" $E$. As one can see,
the relic density is weakly dependent on these details, making our prediction for the relic density also applicable to other fourth lepton family models with suppressed couplings generated by an inverse seesaw-like mechanism. This is especially true, if the WIMP mass is smaller than the Higgs one. Indeed in this case all the contour lines occupy the same ($m_2,\sin\theta$) parameter space. 

Finally let us mention, that usually in the MWT or in other TC models, other new heavy resonances might exist. These could in principle have an impact on our WIMP annihilation cross section, and thus on the $\sin \theta$ slope related to the correct relic density, especially in the heavy WIMP mass region ($m_2 \gtrsim 0.5$ TeV). Indeed, as the new strongly interacting theories predict a spectrum of new states, some of them can act \eg like a new Z' boson. These states could affect the WIMP annihilations near the resonant region \ie when the mediator is nearly on shell $m_2 \approx m_{Z'}/2$, and as long as the mixings and couplings of this new resonance with the SM particles are of the relevant strength. However, in this work we only consider the lightest resonance \ie the Higgs, since other resonances in TC are expected to be heavier ($\sim 1-3$ TeV) and their interactions/couplings to SM particles can be suppressed. This set up is also the most model independent.

\subsection{Collider constraints}
\label{sec:collider}

Here we impose collider constraints on the model using the LEP and the new LHC data. These include the invisible $Z$ boson decay width, the invisible Higgs decay branching fraction, and the mono-$Z$ constraints.
Since these bounds only depend on $\sin\theta$ and $m_2$, they can also be used in other models in which a fourth lepton family is embedded. 

\subsubsection{$Z$-boson decay width }
For the WIMP mass region $m_2 \leq m_Z / 2$, the oblique parameter constraints are replaced by the $Z$ boson decay width constraint in our analysis. As the SM model prediction for the $Z$ boson decay width is in excellent agreement with the measured value, the new Beyond SM contributions to the decay width are highly constrained. As our model could affect the invisible decay channel of the $Z$, the data can be used to constrain our model.
This constraint has been set for this model previously in \cite{Kouvaris:2007iq,Kainulainen:2009rb} and we update this limit here. The Z-decay width constraint reads
\begin{equation}
\sin^4 \theta \left(1-\frac{4m^2_2}{m^2_Z} \right)^\frac{3}{2}  < 0.008 \, ,
\end{equation}
where we have used the uncertainty $\delta N = 0.008$ in the number of light neutrino species $N = 2.984 \pm 0.008 $ reported in PDG \cite{PDG2012} when deriving the constraint\footnote{In the earlier works \cite{Kouvaris:2007iq,Kainulainen:2009rb}, we used $N = 3.00 \pm 0.08$ reported at PDG. 
In this paper we use the more constrained value $N = 2.984 \pm 0.008$ also reported at PDG.}. In Fig.~\ref{fig:Results}, the shaded orange region is disfavored by this bound.
Finally, to avoid the collider limits, set for new charged particles, we demand throughout our analysis that $m_E \geq 500 $ GeV.

\subsubsection{Invisible Higgs decay} 
We can also constraint the model parameter space by using the limits set for the decay branching fractions of the Higgs boson. Indeed, the Higgs boson branching fraction to the invisible sector is constrained to be $R_{ \rm inv.} \leq 0.24$ at 95$\%$ confidence level by the latest LHC data \cite{Giardino:2013bma}. 
The tree-level Higgs decay width to our DM particles is \cite{Heikinheimo:2012yd}
\begin{equation}
\Gamma_{h,N_2}=\frac{G_F m^2_2(C^h_{22})^2 m_h}{2\pi\sqrt2}\left(1-\frac{4m^2_2}{m^2_h} \right)^\frac{3}{2} \, .
\end{equation}
From this, the Higgs boson branching fraction into the invisible sector follows directly $R_{\rm inv. \, N_2} = \Gamma_{h,N_2} /(\Gamma_{h,N_2} + \Gamma_{h,\rm{SM}})$. Here $ \Gamma_{h,\rm{SM}}$ is the total Higgs decay width to the SM particles. Using the equations for $R_{\rm inv. \, N2}$, $ \Gamma_{h,N_2}$ from above, and the limit for $R_{ \rm inv.} \leq 0.24$, we can set a limit for $\sin \theta$ for each fixed DM mass value. In Fig.~\ref{fig:Results}, the shaded red region is excluded by the invisible Higgs decay branching fraction constraint.

Let us also shortly comment about the Higgs phenomenology in the MWT framework \eg the nature of processes like Higgs production $H \leftrightarrow  g g$ and decay $H \rightarrow  \gamma \gamma$ in MWT.  Although these processes are not of course directly related to the fourth heavy neutrino DM scenario, they could potentially be used to test, whether the MWT composite Higgs is favored or not by the LHC data. It turns out, that MWT composite Higgs is consistent with the LHC Higgs data, as was pointed out in~\cite{Belyaev:2013ida}. 

\subsubsection{Mono-Z }
\begin{figure}
\begin{center}
\includegraphics[width=0.5\textwidth]{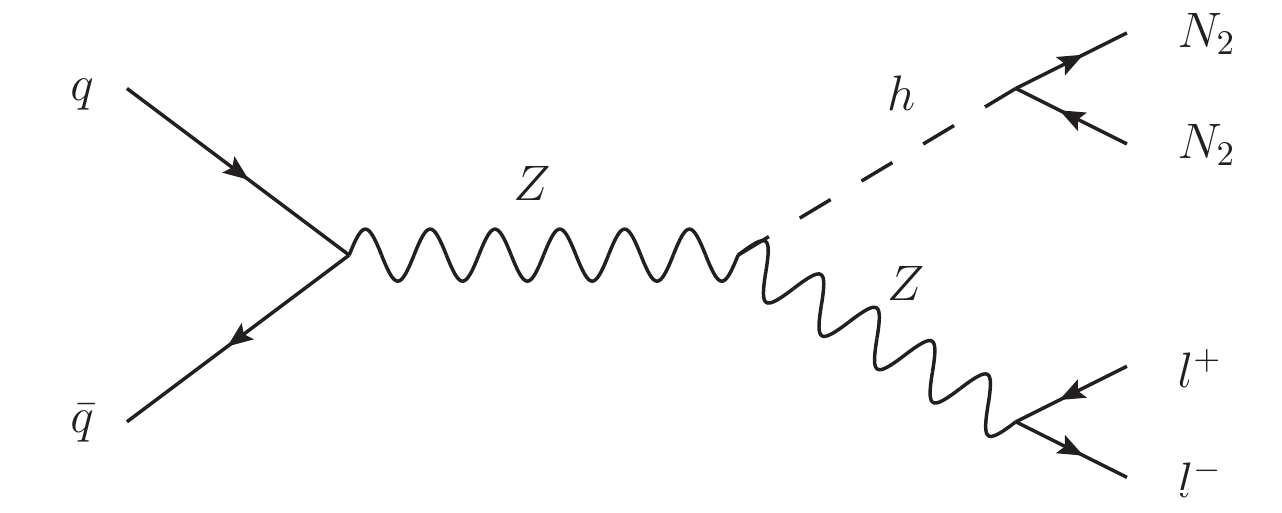}
\caption{\em The most important signal process yielding the mono-Z signature.}
\label{monoZgraph}
\end{center}
\end{figure}

Naturally our WIMP, like any other DM candidate, does not interact directly with the LHC detectors. Nonetheless, if produced, it manifests itself as an excess of missing energy compared to the SM, offering a way to constraint the parameter space of the model. Following Ref.~\cite{Carpenter:2012rg}, we employ a channel in which a $N_2$ pair is produced in association with a $Z$ boson decaying to charged leptons. Notice, that the effective operator approach used in Ref.~\cite{Carpenter:2012rg} breaks down in our case, because the mediators,  \ie the $Z$ or the Higgs bosons, are not heavy enough to be integrated out.  Another important feature of the model is that the Higgs boson can decay to a WIMP pair with a significant branching fraction. Thus the most important channel, yielding the mono-$Z$ signature, is the one in which the Higgs boson is radiated from the $Z$ boson (see Fig. \ref{monoZgraph}). The SM background does not possess a corresponding channel because the Higgs boson  interacts negligibly (or 
not at all) with the SM neutrinos.

Because the effective operator approach is not applicable, we cannot use the existing mono-jet and mono-photon analyses. Based on the analysis of \cite{Shoemaker:2011vi} we do not expect to exclude any part of the parameter space using these channels. We have confirmed this by comparing the parton level mono-jet and mono-photon cross sections against the data shown in  \cite{CMS:rwa} and \cite{Aad:2012fw} respectively. This also means that the mono-Z channel is more constraining, in contrast to the effective operator case, as shown in Ref.~\cite{Carpenter:2012rg}. 

In Ref.~\cite{Carpenter:2012rg} the measurement of $ZZ \to ll\nu\nu$ \cite{Aad:2012awa} with $\sqrt{s}=7$ TeV and integrated luminosity of $4.6$ fb$^{-1}$ by the ATLAS Collaboration is used to derive limits on the DM production. Using the detector acceptance given in \cite{Aad:2012awa}, we calculate the fiducial acceptance. The following cuts define the fiducial region:

\begin{itemize}
\item
two same-flavor opposite-sign electrons or muons, each with $p_{\perp}^l > 20$ GeV, $|\eta^l| < 2.5$;
\item
dilepton invariant mass $76$ GeV $< m_{ll} < 106$ GeV;
\item
no jets with $p_{\perp}^j > 25$ GeV and $|\eta^j| < 4.5$;
\item
$|E_\perp^{\text{miss}}-p_\perp^Z|/p_\perp^Z < 0.4;$
\item
$-\vec{E}_\perp^{\text{miss}}\cdot \vec{p}^Z/p_\perp^Z > 75$ GeV.
\end{itemize}

The leading order simulations are carried out using \texttt{MadGraph 5} \cite{Alwall:2011uj}. The model is implemented into \texttt{MadGraph 5} using the \texttt{FeynRules} \cite{Christensen:2008py} package. The parton level events are passed through \texttt{Pythia 8} \cite{Sjostrand:2007gs} for showering and hadronization, and the jet clustering is done with \texttt{FastJet 3} \cite{Cacciari:2011ma}. The expected number of SM events and the observed number of events, given in \cite{Aad:2012awa}, are
$86.2 \pm 7.2$ and 87 respectively. Using the modified frequentists CLs method \cite{Junk:1999kv,Read:2002hq} we can derive the 90$\%$ confidence level upper limit on the fiducial cross section
\begin{equation}
\sigma_{\text{fid}}(90\% \text{\,CL}) < 5.8 \text{ fb}.
\end{equation}
The excluded parameter space is shown in Fig.~\ref{fig:Results} as a shaded green region. 

\subsection{Direct detection constraints}
\label{sec:direct}

Direct DM searches aim at detecting the nuclear recoil from a scattering between a DM particle and a target nucleus in underground experiments. In our framework the  {\em spin-dependent} (SD)  WIMP-nucleon interaction is $Z$-boson mediated, while the  {\em spin-independent} (SI) is mediated via Higgs boson exchange. The WIMP-nucleon cross sections $\sigma_i^N(\theta, m_2)$ where $i$ stands for the SI or SD parts of the interaction and $N$ refers to a particular nucleon are given in \cite{Kainulainen:2009rb}. 
The SI cross section (in the zero momentum transfer limit) is
\begin{equation}
 \sigma_{\rm SI}^{N} =  (C_{22}^h)^2 \,\frac{8 G_{\rm F}^2\mu_{ N}^2}{\pi} \frac{m_2^2 m_{ N}^2}{m_h^4} f^2,
\label{sigma_SI}
\end{equation}
where  $\mu_N$ is the DM-nucleon reduced mass and $f=\sum_{q = u, d, s} f_{Tq}^{(N)} + 2/9 f_{TG}^{(N)}$ is the Higgs nucleon coupling factor accounting for the quark scalar currents in the nucleons. Here, the model dependent coupling is $C_{22}^h=\sin^2\theta$. 
The SD cross section (in the zero momentum transfer limit) is 
\begin{equation}
 \sigma_{\rm SD}^{N} =  \sin^4\theta \,\frac{8 G_{\rm F}^2\mu_{ N}^2}{\pi} \frac34 a_{ N}^2 ,
\label{sigma_SD}
\end{equation}
where $a_N=\sum_q Y_q \, \Delta_q^{(N)}$ with $Y_q=1/2$ for $q=(u,c,t)$ and $Y_q=-1/2$ for $q=(d,s,b)$.  The values of the scalar and axial-vector couplings with nucleons are taken from \cite{Cheng:2012qr}\footnote{Notice that these coupling factors have relatively large uncertainties. For instance the uncertainty in the scalar coupling $f$ can be as large as a factor of two, following from the $\sigma_{\pi N}$-term and the strange quark content within nucleons that are estimated using lattice techniques (see \eg \cite{Bali:2012qs}) and chiral perturbation theory (see \eg \cite{Cirigliano:2012pq}).}. Having a disposal Eqs.~(\ref{sigma_SI},\ref{sigma_SD}), the  differential rate for DM scattering off a specific isotope is given by 
\begin{equation}\label{RT}
\frac{\ud R_T}{\ud \ER} = \frac{\xi_T}{m_T} \frac{\rho_\odot}{m_2} \sum_{i=\rm SI, SD}   \sum_{N,N'=p,n} \frac{m_T}{2 \mu_N^2} \sigma_i^N(\theta,m_2)\,F_i^{(N,N')}(2 m_T \ER) \int_{v_{\rm min}(\ER)}^{v_{\rm esc}} \hspace{-.60cm} {\rm d}^3 v\frac{ f_\text{E}(\vec v)}v \ , 
\end{equation}
where $\xi_T$ are mass fractions of different nuclides\footnote{$\xi_{T} = 10^3 N_\text{A} m_{T} \zeta_{T} / {\rm kg} \, \bar{A}$, where $N_\text{A} = 6.022 \times 10^{23}$ is Avogadro's number, $\zeta_{T}$ are the numeric abundances and $\bar{A} \equiv \sum_{T} \zeta_{T} A_{T}$.}. Here $v_{\rm min} = \left(m_T \ER / 2 \mu_T^2\right)^{1/2}$ is the minimal velocity providing a given nuclear recoil $\ER$ in the detector, $\rho_\odot=0.3$ GeV/cm$^3$ is the DM energy density at the location of the Earth and $f_\text{E}(\vec v)$ is the DM velocity distribution in the Earth's frame. In this work a customary Maxwell-Boltzmann distribution with velocity dispersion $v_0 = 220$ km/s truncated at $v_{\rm esc} = 544$ km/s  is used \cite{Smith:2006ym}. The nuclear form factors $F_i$ for both SI and SD scattering, which accounts for the non-relativistic physics of the DM-nucleus interaction, are provided in \cite{Fitzpatrick:2012ix}.

For a given  DM mass,  $\theta$ then controls the differential rate and therefore the expected number of events in a given detector. Constraints on $\theta$ have been imposed previously using the \CDMS~\cite{Kouvaris:2007iq}, and   \XENONten-100 results \cite{Kainulainen:2009rb,Heikinheimo:2012yd}. Especially  the experiments based on liquid/gaseous xenon are excellent for the detection of WIMPs with SI interactions due to the large mass of the xenon nuclei. Moreover these detectors are also sensitive to SD WIMP-neutron interactions thanks to the unpaired neutron of the $^{129}$Xe and $^{131}$Xe isotopes. Here we perform a full combined statistical analysis based on the latest result of  \XENON\ and the very recent result of \LUX. 
We have checked that the bound on $\theta$, coming from these experiments,  are the most stringent among those in direct searches. 

\subsubsection{XENON100 } The \XENON\ detector is a two-phase time projection chamber enclosing $62$ kg of active target mass. In \cite{Aprile:2012nq} the collaboration reported a blind analysis with an exposure $w = 34 \times 224.6$ kg$\, \cdot \,$days which yielded no evidence for DM interactions. They found two DM events ($N^{\rm exp}=2$) in the $6.6 - 43.3$ keV$_\text{nr}$ pre-defined nuclear recoil energy window with a background expectation of $N^{\rm bkg} = 1.0 \pm 0.2$ events.


In order to properly reproduce the experimental recoil rate and therefore the predicted number of events in the \XENON\ detector $N^{\rm th}$, one has to convolve Eq.~\eqref{RT} with all the experimental effects; namely the energy resolution of the detector, the detection efficiency and the cuts acceptance. This can be derived by using the formalism of \cite{Aprile:2011hx} and the scintillation efficiency in liquid xenon provided in Fig.~1 of \cite{Aprile:2011hi}.

\subsubsection{LUX}

Like \XENON\, the \LUX\ experiment is a dual-phase xenon time-projection chamber located at the Sanford Underground Research Facility in South Dakota. In \Ref{Akerib:2013tjd}, a non-blind analysis  with an exposure $w$ of $85.3$ live-days $\times$ $118.3$ kg was presented by the collaboration. Since  the collaboration has not still provided detailed information on the expected background and event distribution in the signal region, we assume that the DM events are distributed in an uniform way below and above  the mean of the nuclear recoil event distribution (solid red line in Fig.~3 and 4 of Ref.~\cite{Akerib:2013tjd}). We  restrict our statistical analysis to the region below the red line, where $N^{\rm bkg}=0.64$ electron recoil are expected while the neutron background is instead negligible. In such region  only one event has been found ($N^{\rm exp}=1$). 

The expected number of events below the mean nuclear recoil energy distribution  is then given by
\beq
N^{\rm th} =w\,\frac{1}{2} \int_{E^{\rm min}}^{E^{\rm max}} \hspace{-.33cm} \ud \ER \, \epsilon(\ER) \sum_T \frac{\ud R_T}{\ud \ER} \ ,
\eeq
where the factor $1/2$, as commented above, accounts for the fact that we are considering only half of the rate. Here $E^{\rm min} = 3$ keV$_\text{nr}$ and $E^{\rm max} = 18$ keV$_\text{nr}$ are the lower and upper nuclear recoil energy thresholds (see page 41 of \Ref{GaitskellTalk}) and the efficiency $\epsilon(\ER)$ is obtained by interpolating the black crosses in Fig.~9 of \Ref{Akerib:2013tjd}.

\bigskip
The bounds on $\theta$ can be then inferred by comparing the theoretically predicted number of counts $N^{\rm th}$ to the measured counts $N^{\rm exp}$ in the detector, taking also into account the predicted background $N^{\rm bkg}$. In order to do this we use a standard Likelihood approach and we construct the following statistical test estimator $\lambda=- 2 \ln \left( \mathcal L( N^\text{exp} \, | \, \theta) / \mathcal L_{\rm bkg} \right)$. Here $\mathcal L( N^\text{exp} \, | \, \theta)$  is the likelihood of detecting the number of observed events and $\mathcal L_{\rm bkg}$ is the background likelihood (\ie without DM contribution). Both likelihoods are distributed according to a Poisson distribution since for the null result experiments the number of observed and background events is very low. The constraints are therefore extracted for a certain value of the quantity $\lambda$ which determines the confidence level (CL) of the exclusion. 

An alternative and completely model independent way to derive bounds in direct DM searches has been for the first time presented in \cite{DelNobile:2013sia}. Without entering the details of this paper, one can quickly check our results following the main steps summarized in Sec.~6 of~\cite{DelNobile:2013sia}. In particular, from the first three steps (1a-1c), we can identify the WIMP-nucleon matrix element  in our model to be $\Mel_N=\mathfrak{c}_1^N \Op^\NR_1+\mathfrak{c}_4^N \Op^\NR_4$, where $\Op^\NR_1 = \unop$ and $\Op^\NR_4 = \vec{s}_\chi \cdot \vec{s}_N$ are the non-relativistic operators that account respectively for the SI and the SD part of the interaction with coefficients
\begin{eqnarray}
\mathfrak{c}_1^N (\theta, m_2) & =& \sqrt 2 \,G_{\rm F}\, f  \,\frac{8 m_N^2 m_2^2}{m_h^2}\, \sin^2\theta \ , \\
\mathfrak{c}_4^N (\theta, m_2) & =& -\sqrt 2 \, G_{\rm F} \, a_N 16 m_N m_2 \, \sin^2\theta \ . 
\end{eqnarray}
Thanks to ready-made scaling functions provided in the \href{http://www.marcocirelli.net/NRopsDD.html}{webpage} of \cite{DelNobile:2013sia}, the bound on the free parameter $\theta$ can be simply obtained by following the last two steps (2a-2b).

 In Fig.~\ref{fig:Results} the regions in the ($m_2, \sin\theta$) parameter space above the blue(dark blue) lines are excluded by  \XENON(\LUX) at 90\% CL. The limit becomes stronger for heavy WIMPs, because above roughly 30 GeV the WIMP-nucleus interaction is SI. Indeed, the dominant scattering occurs through Higgs boson exchange with a cross section proportional to $m_2^2$ (see Eq.~\ref{sigma_SI}).

\subsection{Indirect detection constraints}
\label{sec:indirect} 

Indirect searches for DM aim at detecting the final stable SM products of DM annihilations or decays in our Galaxy. This includes charged particles ($e^+$ and $e^-$, $p$ and $\bar p$, deuterium and anti-deuterium), photons (synchrotron radiation, X-rays, $\gamma$-rays) and neutrinos (see \eg \cite{Cirelli:2012tf} for a review of all these signals). 
This is a promising area of research because of many experiments that are currently taking data with different detection techniques. In this work we focus exclusively on bounds coming from neutrino telescope experiments such as \IceCube\ and from $\gamma$-rays satellites, like \FERMI, because they  are already able to probe large portions of the DM parameter space in a wide range of DM masses and primary annihilations channels.

Before moving on, three important remarks, concerning the size of the annihilation cross section and the relevant annihilation final products in our model, are in order; $i)$ depending on the WIMP mass $m_2$, there are two different regimes. If $m_2<m_Z$ the DM candidate mainly annihilates in $b \bar b$ with a cross section $\langle \sigma v \rangle_{\chi \chi \, \rightarrow \, b \bar b}\simeq 2 \times 10^{-27}$ cm$^3$/s (for a reference value $\sin\theta = 0.65$). The cross section is small because 
this channel proceeds mostly in $p$-wave which is naturally suppressed at the present galaxy environments. 
Since the current constraints are not able to probe such small cross sections yet, we will not consider at all this limit in our analysis. On the other hand if $m_2>m_Z$, it is the  $ZZ$, $Z h$ and $t \bar{t}$ channels that dominate the DM annihilation in our Galaxy. These channels include also a significant $s$-wave part, which make the cross sections larger.  In this case, since we are dealing with quite large annihilation cross section $\langle \sigma v \rangle_{\rm tot}\simeq 5 \times 10^{-24}$ cm$^3$/s (for a reference value $\sin\theta = 0.65$), the bounds from indirect searches should be particularly significant in constraining $\theta$;  $ii)$ unlike in the WIMP freeze-out epoch, the annihilation cross section into $W^+W^-$ is negligible at present, being $p$-wave suppressed. Therefore we do not consider at all this channel in our analysis.
$iii)$ the annihilation cross section and thus the branching ratios in specific primary channels are slightly affected by $m_1$  and $m_E$.
For each $m_2$ we have fixed $m_1$  and $m_E$ to those values that satisfy the EW precision data where $\sin \theta$ gives the right relic density. 

Finally, let us comment about the possible relevance of three-body processes in the WIMP annihilation cross section. Three-body final state processes, like emission of gauge boson addition to light fermion-antifermion pair, can have a significant impact on the annihilation cross section in certain cases (see e.g.~related recent works \cite{Yaguna:2010hn,Ciafaloni:2010ti,Ciafaloni:2011sa,Bell:2010ei}). For Majorana WIMP the emission of 'extra' boson in the final state can open up s-wave part to the cross section. Thus, even though the process is higher order, suppressed with extra coupling and propagator factors, in principle it can become of the same order of magnitude, or even larger, as the helicity/velocity suppressed tree level cross section. However, in our case for WIMP masses $m_2 > m_Z$, for which this effect could be somewhat significant, our WIMP already has a large s-wave tree level cross section ($ZZ, Zh$ and $t \bar{t}$ final states), and thus the three-body effect is expected to be subleading (comparable to p-wave part).

\subsubsection{IceCube }

To be captured by the Sun, the WIMP needs to have relatively strong coupling with ordinary matter. As the Sun is mostly made of Hydrogen,  \ie protons carrying spin, the SD WIMP-proton interactions dominate the WIMP capture process. Although SI interactions of WIMPs with heavier elements in the Sun are enhanced by a coherence factor that scales as $\sim A^2$ ($A$ being the total number of nucleons in the nucleus), heavy elements are significantly less abundant. If the WIMP capture rate is large enough, an equilibrium between the WIMP capture and the WIMP annihilation rates in the Sun can take place within the age of the solar system. For the parameter values of our model that lead to a WIMP-nucleus cross section sufficient for establishing such an equilibrium, we can estimate also the annihilation cross section and the neutrino production that is constrained by the neutrino telescopes. For the parameter space where the equilibrium is not established, the WIMP annihilation rate is usually insufficient to 
produce an observable flux of neutrinos. The first estimate for the neutrino production from the annihilation of this DM candidate in the Sun was done in~\cite{Belotsky:2008vh} where there were constraints imposed on the model based on the SuperKamiokande data. Here we use Eqs.~(1-7) of Ref.~\cite{Hooper:2003ui}, (see also \cite{Gould:1987ir,Gould:1991hxnew,Jungman:1995df}), to  calculate the WIMP capture-annihilation rate equilibrium conditions. The equilibrium is achieved, if the relation $t_{\odot} / \tau \gg 1$ is fulfilled. Here 
$t_{\odot} = 4.5 \times 10^9$ years is the age of the Solar system and $\tau = 1/ \sqrt{C_{\odot} A_{\odot}}$ characterizes the time scale at which the equilibrium can be achieved. The quantity $C_{\odot}$ is the WIMP capture rate and the factor $A_{\odot} \propto \langle \sigma v \rangle$ is related to the WIMP annihilation rate (see 
\eg \cite{Hooper:2003ui,Wikstrom:2009kw}). We have used the solar core temperature $T=T_{\odot} \simeq 1.3$ keV for WIMPs,  when calculating the thermally averaged WIMP annihilation cross section $\langle \sigma v \rangle$ in the Sun core. 
With the exception of the points near the two resonances (dips), we found that in the whole WIMP mass region shown in Fig.~\ref{fig:Results},  $t_{\odot} / \tau > 1$, indicating that the equilibrium has been established. 

The \IceCube\ has only presented limits on $b \bar{b}$ (soft) and on $W^+ W^-$ (hard) primary channels \cite{Aartsen:2012kia}. Since however for $m_2>m_Z$, the relevant annihilation products  ($ZZ$, $Z h$ and $t \bar{t}$) generate a hard spectrum of neutrinos, it is a good approximation to assume that the total flux of muons in our model, induced by up-going neutrinos scattered off  ice or more importantly off the rock below the detector, is very similar to the one produced by WIMPs annihilating into $W^+W^-$ (see  \eg Fig.~10 of \cite{Cirelli:2005gh}). One can use the \IceCube\ data to constrain the WIMP annihilation taking place in the Sun to the channels mentioned above. Eventually this constraint can be translated to a constraint on the WIMP-nucleon cross section that dictates the rate of capture of WIMPs in the Sun. As we mentioned above if equilibrium is established the WIMP capture rate  must be equal to the annihilation rate and therefore a constraint on the WIMP annihilation rate in the Sun can be 
interpreted as a constraint on the WIMP-nucleon cross section.
In view of that we can compare the experimental bounds on both $\sigma_{\rm SI}$ and $\sigma_{\rm SD}$ provided in Fig.~2 of \cite{Aartsen:2012kia}  for the $W^+W^-$ primary channel to Eqs.~(\ref{sigma_SI},\ref{sigma_SD}) multiplied by the relevant annihilation branching ratio in our model. The \IceCube\ constraints on the WIMP-nucleon cross section are parametrically equally strict for SI and SD interactions. This is because the Sun is made mostly of hydrogen and helium and therefore the $A^2$ enhancement due to coherence for the SI cross section is not significant. For $m_2<m_Z$, since the primary annihilation channel is $b \bar b$ producing a soft spectrum of neutrinos,  the flux of up-going muons in the detector is several order of magnitude suppressed compared to the $m_2>m_Z$ case and consequently no constraints can be imposed on  $\theta$.  In Fig.~\ref{fig:Results} the shaded magenta areas above the solid ($m_1=0.5$ TeV), dashed ($m_1=1$ TeV) and dotted ($m_1=1.5$ TeV) lines are excluded with 90$\%$ 
confidence level. As one can see the bound is basically independent of $m_1$, $m_E$ and  the details of the underlying theory in which a fourth lepton family is embedded.
Indeed, once WIMPs are captured in the Sun and thermal equilibrium is achieved, the relevant quantities that set the bound are the scattering cross sections showed in Sec.~\ref{sec:direct}.
 
\subsubsection{Fermi-LAT } 
The DM constraints provided by the \FERMI\ $\gamma$-ray data are particularly relevant. For low DM mass (below  30 GeV) and for a variety of primary annihilation channels, they exclude (s-wave) thermally produced DM.

\begin{figure}[t]
\begin{centering}
\includegraphics[width=0.75\textwidth]{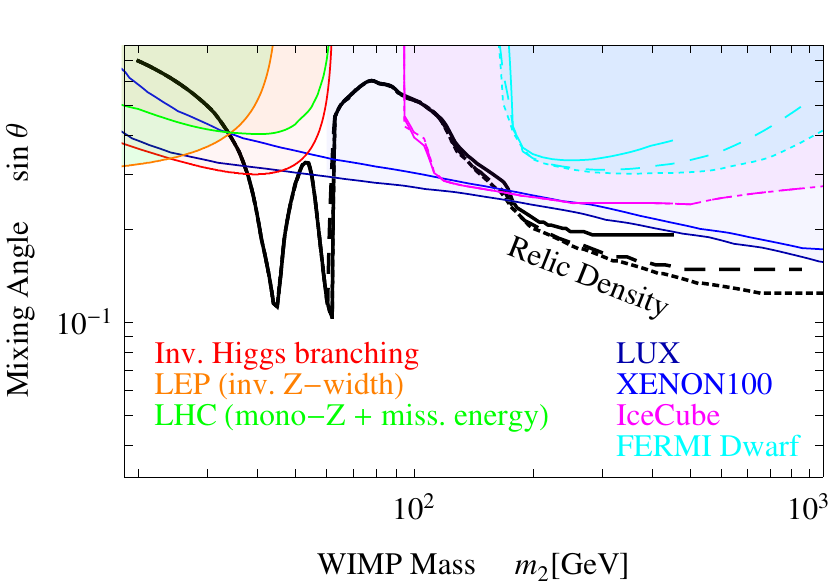}
\caption{\em Parameter space in the ($m_2,\sin\theta$)-plane. The lines along where the DM relic density can be obtained thermally are depicted for three values of $m_1$:  
 black solid ($m_1$=0.5 TeV), black dashed ($m_1$=1 TeV) and black dotted ($m_1$=1.5 TeV). The shaded regions are excluded by the constraints discussed in: Sec.~\ref{sec:collider} (red and orange are set respectively by LHC and LEP measurements of the invisible Higgs branching ratio and Z decay width,  and the green is the mono-Z + missing energy constraints from the latest LHC data),  Sec.~\ref{sec:direct} (blue(dark blue) are the constraints from \XENON(\LUX)), and Sec.~\ref{sec:indirect} (solid ($m_1=0.5$ TeV), dashed ($m_1=1$ TeV) and dotted ($m_1=1.5$ TeV) magenta(cyan) are  set by \IceCube (\FERMI\, dwarf)).}
\label{fig:Results}
\end{centering}
\end{figure}

In particular dwarf satellite galaxies of the Milky Way, due to their large dynamical mass to light ratio and small expected astrophysical background, are among the most interesting targets for DM searches in $\gamma$-rays. Stringent upper bounds on the DM annihilation cross section have been derived from a joint likelihood analysis of 10 satellite galaxies with 2 years of \FERMI\ data \cite{Ackermann:2011wa}. 
The limits are particularly strong for hadronic primary channels and therefore, since in our case the relevant annihilation products are of this class, we expect that such bounds may have an impact on constraining $\theta$. In particular, for $m_2>m_Z$ we rescale the results in Fig.~2 of \cite{Ackermann:2011wa} making the reasonable assumption that the primary products $ZZ$, $Zh$ and $t \bar t$ generate a flux of prompt $\gamma$-rays similar to the one of $W^+W^-$ for the former two channels and to $b \bar b$ for the latter. This is a very good approximation as one can see in Fig.~2 of \cite{Cirelli:2010xx} where the prompt $\gamma$-rays spectra for different annihilation channels are shown. For $m_2<m_Z$,  the annihilation cross section to $b \bar b$ is suppressed by orders of magnitude with respect to the previous case, and consequently $\theta$ is not constrained.  In Fig.~\ref{fig:Results} the shaded cyan regions above the solid ($m_1=0.5$ TeV), dashed ($m_1=1$ TeV) and dotted ($m_1=1.5$ TeV) lines are 
excluded at 95$\%$ confidence level. As one can see for $m_2\gtrsim m_h$  there is a residual dependence on $m_1$ because for heavy WIMP masses, the processes that involve $N_1$ as a mediator affect somewhat the amplitude of the annihilation cross section.

Other important and strong limits on the annihilation cross section for different primary annihilation channels have been set by the $\gamma$-ray diffuse emission measurement by \FERMI\ at intermediate latitudes \cite{Cirelli:2009dv,Papucci:2009gd,Ackermann:2012rg,Zaharijas:2010ca}. In particular, the most recent limits come from 2 years of observations in the region $5^{\circ}\leq b \leq15 ^{\circ},~-80^{\circ}\leq \ell \leq 80^{\circ}$, where $b$ and $\ell$ denote the galactic latitude and longitude respectivelty \cite{Zaharijas:2010ca}. However, in the WIMP mass region we are interested in, the limits are weaker compared to the ones coming from dwarf galaxies, and therefore we do not include them in our analysis. 

\section{Results and conclusions}
\label{sec:results}
Our main results are presented in Fig.~\ref{fig:Results} in the $(m_2, \sin \theta)$ parameter space of the theory. The black lines represent the parameter space where the relic density of the particle $N_2$ is that of DM and the model passes the EW precision tests. The three interpolating black lines  \ie solid, dashed and dotted represent three different values of $m_1=0.5,~1,~1.5$ TeV respectively. For every point on the black line, the mass $m_E$ is fixed to a value that makes the model pass the EW precision tests. 
 As we explained in Sec.~\ref{sec:omega} the two dips in the $\sin \theta$ are taking place approximately at half the mass of $Z$ and Higgs bosons reflecting the fact that at these mass scales the huge increase of the annihilation cross section due to resonance is compensated by tuning down the value of $\sin \theta$.  As it can be seen, up to $m_2 \sim m_W$ the three lines are identical. This is because before the onset of the $W^+W^-$ ($Z Z$) annihilation channel, $m_E$ ($m_1$) is irrelevant to the annihilation cross section.

We also show all the relevant constraints in the figure. Constraints on the width of the invisible $Z$ decay in LEP, the mono-$Z$ constraint from LHC and the constraint from the invisible  Higgs branching exclude the low mass phase space with $m_2 \lesssim 38$ GeV. Direct detection constraints coming from experiments based on liquid/gaseous xenon (\XENON\ and \LUX) exclude the mass range of $m_2$ from $\sim 62$ GeV to $\sim 188$ GeV. Therefore the overall allowed region of the model is $ 38~{\rm GeV}\lesssim m_2 \lesssim  62 ~{\rm GeV}$ and $m_2 \gtrsim 188 ~{\rm GeV}$ (for $m_1 = 500$ GeV). 


In conclusion, in this paper we presented the current limits from all possible constraints on heavy neutrinos with helicity suppressed couplings as thermally produced dark matter. Heavy neutrinos can easily emerge from a fourth lepton family. Although severe constraints exist on the existence of a fourth quark family, a new lepton family can emerge easily from theories beyond the SM. We showed as an example the embedment of a new lepton family in the context of TC. We identified the parameter space that produces the DM relic abundance and passes the EW precision tests. In addition we imposed constraints from colliders; the invisible $Z$ decay from LEP, and the invisible Higgs decay from LHC. Moreover we studied the mono-$Z$, mono-jet and mono-photon constraints arising from LHC. Because  
the mediators of the heavy neutrinos are the $Z$ and Higgs bosons, the CMS constraints based on non-detection of excessive missing energy are not applicable. This is due to the fact that the 
CMS results are valid upon the assumption that the mediators are heavy and therefore WIMPs have contact interactions with the partons. 
Furthermore, we updated the limits on the heavy neutrinos imposed by underground direct search experiments based on liquid/gaseous xenon. Finally, we also studied possible limits following from indirect DM detection. In particularly we set constraints for our model using the \IceCube\  and \FERMI\ $\gamma$-ray data. We find that heavy neutrinos can play the role of thermally produced DM within the mass ranges
\begin{equation} 
38~{\rm GeV}\lesssim m_2 \lesssim  62 ~{\rm GeV}\qquad \mathrm{and} \qquad m_2 \gtrsim 188 ~{\rm GeV}.
\end{equation}
The second limit becomes sligthly weaker $m_2 \gtrsim 182 ~{\rm GeV}$ for $m_1 = 1500$ GeV.
We should emphasize that although we chose to embed our fourth lepton family in a TC framework, our derived results and constraints 
are very little model dependent and thus 
applicable generically to any model where heavy neutrinos with suppressed couplings play the role of thermally produced DM. 

{\small 
\paragraph{Acknowledgments.}

\noindent We would like to thank Marco Cirelli, Mads Frandsen and Marco Nardecchia for useful discussions. The CP3-Origins centre is partially funded by the Danish National Research Foundation, grant number DNRF90. The work of T.H. was supported by the European Commission through the ``LHCPhenoNet'' Initial Training Network PITN-GA-2010-264564. The work of M.J. was supported in part by European Union's Seventh Framework Programme under grant agreements (FP7-REGPOT-2012-2013-1) no 316165, PIF-GA-2011-300984, the EU program ``Thales'' MIS 375734  and was also co-financed by the European Union (European Social Fund, ESF) and Greek national funds through the Operational Program ``Education and Lifelong Learning'' of the National Strategic Reference Framework (NSRF) under ``Funding of proposals that have received a positive evaluation in the 3rd and 4th Call of ERC Grant Schemes''.}

\end{document}